\documentclass[12pt]{JHEP3}

\usepackage{epsfig}
\usepackage{amsmath}
\usepackage{amssymb,amsfonts}
\usepackage{graphicx}
\usepackage{cite}

\makeatletter
\renewcommand{\@makecaption}[2]{%
{ #2}%
}

\makeatother

\newcommand{\ie}{{\em i.e. }}

\newcommand{\R}{\mathbb{R}}
\newcommand{\Co}{\mathbb{C}}

\renewcommand{\P}{\mathbb{P}^1}

\newcommand{\Z}{\mathbb{Z}}

\newcommand{\C}{{\cal C}}
\newcommand{\G}{{\cal G}}

\newcommand{\N}{{\cal N}}

\newcommand{\bZ}{\bar{Z}}
\newcommand{\si}{\sigma}
\newcommand{\tpsi}{\tilde{\psi}}

\newcommand\fverb{\setbox\pippobox=\hbox\bgroup\verb}
\newcommand\fverbdo{\egroup\medskip\noindent%
                        \fbox{\unhbox\pippobox}\ }
\newcommand\fverbit{\egroup\item[\fbox{\unhbox\pippobox}]}
\newbox\pippobox

\title{On the equivalence of $\N=1$ brane worlds and geometric singularities
with flux}

\author{P. Kaste$^1$ and H. Partouche$^2$ \\

$^1$ Institute for Theoretical Physics\\
ETH H\"onggerberg\\
CH--8093 Z\"urich, Switzerland \\
 E-mail: \email{kaste@itp.phys.ethz.ch}\\
\vskip .2cm

$^2$ Centre de Physique Th{\'e}orique, Ecole Polytechnique, \\
F-91128 Palaiseau cedex, FRANCE\\
E-mail: \email{Herve.Partouche@cpht.polytechnique.fr}\\
}
\preprint{\hepth{0409303}\\ CPHT-RR-050.0804}

\abstract{We consider Kaluza Klein reductions of M-theory on the
$\Z_N$ orbifold of the spin bundle over $S^3$ along two different $U(1)$
isometries. The first one gives rise to the familiar ``large $N$
duality'' 
of the $\N=1$ $SU(N)$ gauge theory in which the UV is realized as the
world-volume theory of $N$ D6-branes wrapped on $S^3$, whereas the IR
involves $N$ units of RR flux through an $S^2$. 
The second reduction gives an equivalent version of this duality in which
the UV is realized geometrically in terms of a $\P$ of
$A_{N-1}$ singularities, with one unit of RR flux  
through the $\P$. The IR is
reached via a geometric transition and involves a single D6 brane on a
lens space $S^3/\Z_N$ or, alternatively,
a singular background $(S^2\times \R^4)/\Z_N$, with one unit of
RR flux through $S^2$ and, localized at the singularities, an action 
of their stabilizer group in the
$U(1)$ RR gauge bundle, so that no massless
twisted states occur. 
We also consider linear $\sigma$-model
descriptions of these backgrounds.}


\begin{document}

\maketitle 

\setlength{\baselineskip}{1.2\baselineskip}

\section{Introduction}

The physical motivations for studying type II string theories increased a lot 
when, in the course of the heterotic/type II dualities
\cite{Hull:1994ys,Witten:1995ex}, it was realized that they as well can
describe non-Abelian gauge theories. For example, it has been shown in
\cite{Hull:1995mz} that such theories, with $\N=4$
supersymmetry in 4 dimensions, arise as type II compactifications on 
$K3\times T^2$ backgrounds, where the $K3$ develops an $A$-$D$-$E$ type 
singularity (see also \cite{Aspinwall:1995zi}). 
Similarly, by considering a compactification on a Calabi-Yau (CY) 3-fold that 
contains a curve of such singularities, the supersymmetry is reduced  to 
$\N=2$, 
\cite{Kachru:1995wm,Ferrara:1995yx,Kachru:1995fv,Antoniadis:1995cy,
Klemm:1996kv}.  
In the type IIA models, the non-Abelian gauge degrees of freedom arise 
nonperturbatively from D2-branes wrapped on the 
vanishing 2-cycles of a singularity that locally looks like  
$\R^4/\Z_N\times S^2$ in the case of the pure $SU(N)$ gauge theory 
\cite{Hull:1995mz}. 
At this stage, a natural strategy to further 
reduce the supersymmetry to $\N=1$ would be to add Ramond-Ramond (RR)
flux on $S^2$. 
This is precisely 
what we want to consider in this paper.

In the meanwhile, however, attention has focused on
another realization of non-Abelian gauge theories
in type II strings, namely the effective world-volume theory
on D-branes generated by open strings ending on them.
$\N=1$ gauge theories in 4 dimensions can be described this way by 
considering the type IIA string on a smooth CY manifold with
3 directions of $N$ coincident D6-branes wrapped on a compact special 
Lagrangian submanifold, while the other 3+1 world volume directions fill 
a transverse $\R^{3,1}$.
A simple local example is given by $N$ D6-branes wrapped on the 
special Lagrangian $S^3$ of the non-compact deformed conifold. 
In the classical large radius limit, in which stringy 
$\alpha^{\prime}$-corrections are suppressed, this 
describes the pure $SU(N)$ super Yang-Mills (SYM) theory 
in the UV, with the gauge coupling constant being related to the 
volume of the $S^3$ roughly as $1/g_{YM}^2\sim {\rm vol}(S^3)$. 
In the quantum parameter space, due to the complexification with the 
$C$-field or $B$-field respectively,
one can follow the system smoothly through a geometrical conifold transition 
in the course of which the $S^3$ shrinks to zero size and gets replaced by
an $S^2$.  
After the transition, the system is 
described by $N$ units of RR flux through the $S^2$   of the resolved 
conifold and is associated to the (confining) IR of the gauge
theory. 
The effect of the flux is to partially break supersymmetry from $\N=2$ to
$\N=1$ \cite{Taylor:1999ii,Mayr:2000hh} by adding a magnetic
Fayet-Iliopoulos term \cite{Antoniadis:1995vb,Partouche:1996yp}
for the closed string $U(1)$ associated to the homology class of the $S^2$.
This $S^3\to S^2$ transition, with branes on the one hand side and flux on
the other, is the prime example of the ``large $N$ duality'' of 
\cite{Vafa:2000wi}. 
As there are two birationally equivalent small resolutions of the 
singular conifold, there are actually two equivalent realizations of the IR 
phase, with their $S^2$'s related by a flop transition.  
In \cite{Atiyah:2000zz}, it was shown how to derive 
the connection between the UV and IR phases from
Kaluza-Klein (KK) reduction  
of M-theory on a background of $G_2$ holonomy, which is
an orbifold of the spin bundle over $S^3$ that we denote by 
$Spin(S^3)/\Z_N$.
The complex one-dimensional quantum parameter space of these models indeed
connects three classical limits \cite{Atiyah:2001qf}: 
One of them associated to the UV and the other two to 
the IR of the $SU(N)$ gauge 
theory\footnote{Another description of these
phases in terms of a D6-brane in $\Co^2/\Z_N\times \Co$ can be found in 
\cite{Aganagic:2001jm}.}.

In this paper, we shall derive and 
complete
an alternative type IIA picture of the same
large $N$ duality, already mentioned in \cite{Giveon:2001uy}, 
by reducing the M-theory background along a different $U(1)$ isometry. 
In this version, the UV of the $SU(N)$ 
SYM theory is realized in type IIA by a $\P$ of $A_{N-1}$ singularities, with 
one unit of RR flux through the $\P$. 
This set up generalizes the spontaneous breaking of $\N=2$ to $\N=1$ of 
\cite{Taylor:1999ii,Mayr:2000hh} to the non-Abelian case of $U(N)$. 
It involves a 
magnetic Fayet-Iliopoulos  term for the diagonal $U(1)$ associated to the 
homology class of the $\P$. The corresponding field theory analysis 
extending  the model of  \cite{Antoniadis:1995vb,Partouche:1996yp} 
has appeared in the recent paper \cite{Fujiwara:2004kc}.
In type IIA, 
the gauge coupling of such models is again related to the
volume of the base $\P$ as $1/g_{YM}^2\sim {\rm vol}(\P)$. 
The IR physics involves after a conifold 
transition  a single D6-brane wrapped on a lens space $S^3/\Z_N$. 
The third branch, also associated to the IR, involves a 
singular background of 
topology $(S^2\times \R^4)/\Z_N$, with one unit of RR flux through
$S^2$, and can be reached 
geometrically from the UV phase by a flop transition. 
For $N$ even, the orbifold group acts as $\Z_2\times\Z_{N/2}$, whose $\Z_2$
subgroup fixes an $S^2/\Z_{N/2}$ of $A_1$ singularities, while for any $N$
the north and south poles of the $S^2$ (times the origin in $\R^4$) are
fixed under the whole $\Z_N$. 
There are, however,
no massless states associated to these singularities since, at their locus,
there is a free action of their stabilizer group 
within the $U(1)$ gauge fiber associated to the 
RR one-form potential. This is the same mechanism that removes massless
states from twisted sectors in the type IIA dual of the six-dimensional
CHL compactification \cite{Schwarz:1995bj}.

The paper is organized as follows.
Section \ref{KKreduc} is devoted to the KK reductions of the three 
classical phases of the  M-theory background along the two different
$U(1)$ isometries. In particular, it clarifies how the original large 
$N$ duality involving $N$ D6-branes is on equal footing with the picture based 
on the $A_{N-1}$ singularity with flux in type IIA. 
For this second point of view, we study in Section \ref{transition} how the 
three phases are connected via conifold or flop transitions. 
Finally, in Section \ref{sigma} we present 
linear $\sigma$-model descriptions of the six-dimensional type IIA  
backgrounds, together with their lifts to seven dimensions associated to the 
$G_2$ holonomy spaces occurring in M-theory.

\section{KK reductions of $G_2$ holonomy spaces to type IIA}
\label{KKreduc}

We start by considering the geometrical realization
of the $\N=1$ pure $SU(N)$ gauge theory  in M-theory 
\cite{Atiyah:2000zz, Acharya:2000gb}. 
The relevant background is the $G_2$ holonomy manifold $Spin(S^3)$ modded by 
$\Z_N$.  Classically, the associated moduli 
space is composed of three branches
\cite{Atiyah:2001qf}.\footnote{The branches for similar models based
on spaces of the form $(\mbox{CY}\times S^1)/\Z_2$ have also been
considered: Examples involving compact CY's are treated in
\cite{Kaste:2001iq}, while the case where the CY is the deformed
conifold is introduced in \cite{Partouche:2000uq}.}  
In one of them, the seven-dimensional 
orbifold is singular and the massless spectrum contains the non-Abelian 
$SU(N)$ vector multiplets. In the last two branches, the $\Z_N$ orbifold 
acts freely and there are no massless non-Abelian gauge bosons anymore. 
In the description 
of \cite{Atiyah:2001qf}, the $SU(2)^3$ isometry of the $G_2$ holonomy
metric on 
$Spin(S^3)$ is explicit and the orbifold group $\Z_N$ is taken to be a 
subgroup of the diagonal $U(1)$ in the first of these $SU(2)$'s. 
The KK reduction along this $U(1)$ subgroup containing the $\Z_N$ was
considered in \cite{Atiyah:2000zz} and 
allowed to associate the moduli space branches to the phases involved in  
the large $N$ duality in type IIA of \cite{Vafa:2000wi}. 
Performing now the KK reduction along another $U(1)$ 
isometry that does not contain the $\Z_N$, 
we derive 
and complete the alternative version of the large $N$ duality introduced in  
\cite{Giveon:2001uy}.

The spin bundle over $S^3$, $Spin(S^3)$, is a 7-dimensional manifold that
is asymptotically a cone over $S^3\times S^3$.
A convenient description of the $S^3 \times S^3$ base is in terms of a coset 
$SU(2)^3/SU(2)_D$, where $SU(2)_D$ is the diagonal $SU(2)$ factor acting on 
the right: 
\begin{equation}
\begin{array}{lcl}
 & & (g_1,g_2,g_3) \in SU(2)^3\, ,\\
\mbox{such that}&&(g_1,g_2,g_3)\equiv (g_1h,g_2h,g_3h)\, ,\,\, h\in SU(2)_D\, .
\label{s3s3}
\end{array}
\end{equation}
The 7-manifold is then constructed by replacing one of the $S^3\simeq SU(2)$ 
factors by $\R^4$, \ie one allows one of the 
$g_i$'s $(i=1,2,3)$ to take values in the set of $2\times 2$ 
matrices of the form 
\begin{equation}
g=\begin{pmatrix}
Z & iZ' \cr i\bar{Z}' & \bar{Z}
\end{pmatrix}\, ,
 \qquad \mbox{where}\quad Z,Z' \in \Co\, ,
\label{r4}
\end{equation}
but without constraining the determinant to be one. 
By choosing which of the three $S^3$ 
factors is replaced, one obtains this way three isomorphic 7-manifolds 
\cite{Atiyah:2001qf} that 
admit asymptotically conical metrics of
G$_2$ holonomy \cite{salom,Gibbons:1989er}. 

With respect to these metrics, there is an 
$SU(2)_{1,L}\times SU(2)_{2,L}\times SU(2)_{3,L}$
group of isometries acting on these manifolds that is
realized by left multiplication,
\begin{equation}
\begin{array}{lcl}
 & & (g_1,g_2,g_3)\to (h_1g_1,h_2g_2,h_3g_3)\, ,\\
\mbox{where}&&(h_1,h_2,h_3)
\in SU(2)_{1,L}\times SU(2)_{2,L}\times SU(2)_{3,L}\, .
\end{array} 
\label{iso1}
\end{equation}
At this stage, the three manifolds are smooth. In order to describe an $SU(N)$ 
gauge theory, one considers a $\Z_N$ orbifold of them. This $\Z_N$ is 
chosen 
as a discrete subgroup of $SU(2)_{1,L}$, its generator being 
\begin{equation}
\xi=
\begin{pmatrix}
e^{2\pi i/N} & 0 \cr 0  & e^{-2\pi i/N}
\end{pmatrix}\, .
\label{zn}
\end{equation}
Note that the left action $\xi g$ amounts to 
$(Z,Z')\to e^{2\pi i/N}(Z,Z')$  on the coordinates in (\ref{r4}), 
whereas a right action $g\xi$ would act as 
$(Z,Z')\to (e^{2\pi i/N}Z,e^{-2\pi i/N}Z')$. 

We now describe the three spaces 
$\G_I$, $\G_{I\!I}$ and $\G_{I\!I\!I}$ obtained this way and reduce them 
along two 
different $U(1)$ isometry groups, the diagonal 
\begin{equation}
U(1)_{1,L}\subset SU(2)_{1,L}\quad \mbox{and} 
\quad U(1)_{2,L}\subset SU(2)_{2,L}\, ,
\label{KKu1}
\end{equation}
which are relevant to derive the dualities of \cite{Vafa:2000wi} and 
\cite{Giveon:2001uy}, 
respectively\footnote{As a consequence
of the fact that asymptotically the $G_2$ metric we consider is
conical, the string coupling in the
associated type IIA models is unbounded. In \cite{Brandhuber:2001yi}, another
metric of $G_2$ holonomy is given that is not asymptotically
conical and has a reduced isometry group $U(1)_{1,L}\times SU(2)_{2,L}\times
SU(2)_{3,L}\times \Z_2$. Considering the orbifold by $\Z_N\subset U(1)_{1,L}$,
the KK reduction along $U(1)_{1,L}$ gives rise to type IIA with 
$N$ D6-branes on $S^3$ with a bounded dilaton, while a reduction along
$U(1)_{2,L}\subset SU(2)_{2,L}$ gives our alternative realization as
a singularity with flux in which, however, the dilaton is unbounded.
}.
Some explicit expressions for the $G_2$ metrics and the relevant isometry 
actions  are given in Appendix \ref{metrics}. 

\vspace{.3cm}
$\bullet$ Fixing the gauge $g_3\equiv 1$ in Eq. (\ref{s3s3}), the space 
$\G_I$ can 
be parametrized  as\footnote{Our convention is to use square brackets when
referring to representatives of equivalence classes.}
\begin{equation}
\G_I \, : \quad \left[\xi^k g_1, g_2, 1\right]\, ,
\quad g_1\in \R^4\, , \quad g_2\in SU(2)\, ,
\label{GI}
\end{equation}
where it is understood that points with different integer values of $k$ are 
identified. 
Topologically, this space is the orbifold $\R^4 /\Z_N \times S^3$ and there 
are massless vector multiplets occurring in M-theory. This describes 
the $SU(N)$ SYM theory in the UV. 
To perform the KK reduction of $\R^4/\Z_N$ along  $U(1)_{1,L}$, think of 
$\R^4/\Z_N$ as the disjoint union of lens spaces $S^3/\Z_N$
of arbitrary radii $\rho\in \R_{\geq 0}$. 
As recalled in Appendix \ref{KKlens},
the Hopf reduction of each such lens space is an $S^2$ of radius $f(\rho)$,
where $f\, :\, \R_{\geq 0}\to \R_{\geq 0}$ is one-to-one and onto, and 
with $N$ units of RR flux through it. Letting $\rho$ vary, one sees that
$(\R^4/\Z_N)/U(1)_{1,L}$ is 
$\R^3$
with a magnetic source of $N$ units for the RR two-form field strength at 
the origin, \ie there are $N$ D6-branes at the vanishing locus of the orbit 
of $U(1)_{1,L}$, \cite{Sen:1997kz}. This type IIA background is 
the deformed conifold\footnote{The metric on this space is of course not the 
Calabi-Yau metric due to the back-reaction to the branes. The same comment
will apply to all conifold spaces in the presence of branes or fluxes that
we shall encounter. The respective metrics can easily be derived by using
the expressions given in Appendix \ref{metrics}.} 
$T^* S^3$ with $N$ D6-branes wrapped on $S^3$. 

In the same spirit, we can consider the KK reduction of $\G_I$ along 
$U(1)_{2,L}$. Since $U(1)_{2,L}$ acts on $S^3\simeq SU(2)$ only, the 
$\R^4/\Z_N$ factor 
remains as it is. Seen as an Hopf fibration, the reduction of $S^3$ gives 
rise to a two-sphere  with one unit of RR flux through $S^2$ 
(see Appendix \ref{KKlens} with $N=1$).  
Topologically, the type IIA background is thus a $\P$ of 
$A_{N-1}$ singularities, $\R^4/\Z_N \times S^2$, that we shall denote $\C_I$. 
It 
is the resolved conifold  with one unit of RR flux on the $S^2$   and a 
$\Z_N$ orbifold action on the $\R^4$ fiber.
\vspace{.3cm}

$\bullet$ Again fixing $g_3\equiv 1$ thanks to  Eq. (\ref{s3s3}), the manifold 
$\G_{I\!I}$ 
can be defined as
\begin{equation}
\G_{I\!I} \, : \quad \left[\xi^k g_1, g_2, 1\right]\, ,
\quad g_1\in SU(2)\, ,\quad  g_2\in \R^4\, ,
\label{GII}
\end{equation}
where an identification of the points with different  values of $k$ is 
understood. There are no fixed points under the orbifold and topologically 
the manifold is $S^3/\Z_N\times \R^4$, where $S^3/\Z_N$ is a lens space.  
There is no massless non-Abelian gauge multiplet as expected for the 
$SU(N)$ SYM theory 
in the IR. The KK reduction along $U(1)_{1,L}$ acts only on the $S^3/\Z_N$ 
factor and 
gives a two-sphere with $N$ units of RR flux through it, as reviewed in 
Appendix \ref{KKlens}. 
Altogether, the topology is thus $S^2 \times \R^4$ and corresponds to
a type IIA  
string on the resolved conifold with $N$ units of RR flux through 
the   $S^2$. 
The presence of flux has the effect to partially break
$\N=2$ to $\N=1$. This effect is a particular case of the general set up of
\cite{Taylor:1999ii,Mayr:2000hh} that embeds in string theory the field theory 
mechanism of \cite{Antoniadis:1995vb,Partouche:1996yp} for rigid
supersymmetry.  

For the reduction of $\G_{I\!I}$ along $U(1)_{2,L}$, only the $\R^4$ factor is 
concerned and gives $\R^3$, with
a single D6-brane at the origin that wraps the lens space. One thus 
obtains 
a background whose topology is  $S^3/\Z_N\times \R^3$, the deformed conifold 
whose base is modded by a freely acting $\Z_N$, with one 
D6-brane wrapped on the lens space. We shall call this space
$\C_{I\!I}$. 
Since the D6-brane carries one unit of magnetic charge w.r.t.\ the type
IIA RR one-form gauge potential, there is conservation of the RR charge between
the present background and $\C_{I}$ with one unit of flux.

\vspace{.3cm}

$\bullet$ Finally, in the gauge $g_2\equiv 1$ in Eq. (\ref{s3s3}), the
manifold  
$\G_{I\!I\!I}$ can be defined as 
\begin{equation}
\G_{I\!I\!I} \, : \quad \left[\xi^k g_1, 1,g_3\right]\, ,
\quad g_1\in SU(2)\, ,\quad g_3\in \R^4\, .
\label{GIII}
\end{equation}
It is isometric  to $\G_{I\!I}$ via the exchange $g_2\leftrightarrow g_3$
and thus associated to the IR of the SYM theory 
as well. The KK reduction along $U(1)_{1,L}$ gives then again the resolved 
conifold with $N$ units of RR flux through the   $\P$. 
The $\P$'s obtained 
by $U(1)_{1,L}$ reduction  of $\G_{I\!I}$ and $\G_{I\!I\!I}$ are related by 
a flop. 

To reduce $\G_{I\!I\!I}$ along $U(1)_{2,L}$, it is convenient to choose 
instead the gauge
$g_1\equiv 1$ in Eq. (\ref{s3s3}), so that we have :
\begin{equation}
\G_{I\!I\!I} \, : \quad \left[1, g_2\xi^{-k},g_3\xi^{-k}\right]\, ,
\quad g_2\in SU(2)\, ,\quad g_3\in \R^4\, ,
\label{GIII2}
\end{equation}
where points with different values for $k$ are identified. In this 
parametrization, the 
freely acting $\Z_N$ acts simultaneously from the 
{\em right}  on the $S^3$ and $\R^4$ factors. 
As shown in Appendix \ref{KKR}, after KK reduction to type IIA, 
the background topology is $(S^2\times \R^4)/\Z_N$, with one unit of
RR flux through $S^2$. 
It corresponds to an orbifold of the resolved conifold we shall call 
$\C_{I\!I\!I}$. 
In order to discuss the orbifold action in type IIA, we distinguish the
cases of odd respectively even $N$, parametrizing the $\R^4$ by two 
complex coordinates $Z_3$ and $Z_3'$ in both cases.
For odd $N$, the orbifold group acts on the two-sphere
as rotations around the axis
through the north and south poles by angles $2\pi k/N$, combined with an
action
\[
(Z_3,Z_3')\equiv\left( (-1)^k e^{i\pi k/N}Z_3,
(-1)^k e^{-i\pi k/N}Z_3'\right)\, 
\]
on $\R^4$, for $k=0,\ldots,N-1$. 
This geometrical orbifold action fixes two points of the six-dimensional
space, namely the north and south poles of the $S^2$ (times the origin in
the $\R^4$-fiber). 
Localized at these two fixed points, however, there
is an embedding of the
$\Z_N$ orbifold action within the $U(1)$ gauge fiber associated with the
RR one-form potential. The orbifold doesn't act on the gauge fibers over
any other non-fixed point. 
For even $N$, the orbifold group acts as $\Z_2\times \Z_{N/2}$, 
where the $\Z_{2}$ acts trivially on the two-sphere 
and the $\Z_{N/2}$ by rotations around the axis
through the north and south poles by angles $4\pi l/N$, whereas the
$\Z_2\times\Z_{N/2}$ acts on the $\R^4$ as
\[
(Z_3,Z_3')\equiv\left( e^{2i\pi (l+\frac{N}{2}\alpha)/N}Z_3,
e^{-2i\pi (l+\frac{N}{2}\alpha)/N}Z_3'\right)
=\left( e^{2i\pi k/N}Z_3,
e^{-2i\pi k/N}Z_3'\right)\, ,
\]
with $k=l+\frac{N}{2}\alpha$, where $l=0,\ldots,\frac{N}{2}-1$ 
and $\alpha=0,1$.
Note that the generator of $\Z_2$ always leads to an 
$S^2/\Z_{N/2}$ of $A_1$ singularities given by $Z_3=Z_3'=0$.
The poles of the base (times the origin in the fiber)
are always fixed by the whole $\Z_N$. 
Again there is a localized embedding of the $\Z_2$ (respectively $\Z_N$) 
orbifold action in 
the $U(1)$ gauge symmetry associated with the RR one-form potential
precisely at the $\Z_2$ (respectively $\Z_N$)-fixed loci of the
geometrical orbifold action.

In a nutshell, the free $\Z_N$ orbifold action on $\G_{I\!I\!I}$
reduces to a geometrical $\Z_N$ orbifold action on $\C_{I\!I\!I}$ 
with $\Z_N$ and $\Z_2$-fixed loci (the latter for even $N$). 
The type IIA theory, however, ``remembers'' that the action on 
$\G_{I\!I\!I}$ was free by a localized free orbifold action of the stabilizer 
subgroups $\Z_N$ and $\Z_2$ on the RR $U(1)$ gauge fiber at the
respective fixed point loci.  
These additional RR gauge twists localized at the fixed point set of 
the geometric orbifold actions are responsible for removing the massless 
states that would otherwise occur through their twisted sectors.
For $N=2$, it is these RR gauge twists that make the difference between the
otherwise identical models $\C_{I}$ and $\C_{I\!I\!I}$, where the former
has an $SU(2)$ gauge symmetry whereas the latter does not. 
This is exactly the same mechanism that reduced the gauge symmetry
in the type IIA dual of the six-dimensional CHL compactification
\cite{Schwarz:1995bj}. As in that case, the precise working of this
mechanism from the type IIA perspective is not well understood, due to the
lack of a CFT description of these nontrivial RR backgrounds.
It looks very much like introducing an additional circle corresponding
to the RR $U(1)$ fiber (the M-theory circle) with an ordinary shift orbifold
action on it.
The result of removing massless states, 
however, can be inferred from the fact that the orbifold action 
is free in the dual M-theory realization $\G_{I\!I\!I}$
due to the shift in the M-theory circle. 

\section{Transitions from the type IIA  point of view}
\label{transition}

The KK reductions of the $G_2$ holonomy spaces $\G_I$, $\G_{I\!I}$ and
$\G_{I\!I\!I}$ along $U(1)_{2,L}$ give rise to the
6-dimensional spaces $\C_I$, $\C_{I\!I}$ and $\C_{I\!I\!I}$, together
with RR flux or
a D6-brane. In this 
section are going to discuss the geometric transitions between the
different type IIA geometries.
 
$\C_I$ is the resolved conifold with a $\Z_N$ modding  action on the fiber. 
Let us introduce four complex variables $z_i$ $(i=1,\ldots,4)$ and homogeneous 
coordinates $[\xi_1,\xi_2]$ for the base $\P$. The resolved conifold can then
be defined as 
\begin{equation}
\left\{
\begin{array}{l}
(z_1+iz_2)\xi_1-(z_3+iz_4)\xi_2=0\\
(z_3-iz_4)\xi_1+(z_1-iz_2)\xi_2=0
\end{array}
\right.\, .
\label{resolved}
\end{equation}
$\C_I$ is obtained under a discrete isometry identification that
we can define as
\begin{equation}
\mbox{\big (} (z_1\pm iz_2), (z_3\pm iz_4)\mbox{\big )}\equiv 
\left( e^{\pm 2\pi i/N}(z_1\pm iz_2),  
e^{\pm 2\pi i/N}(z_3\pm iz_4) \right)\, .
\label{ZN1}
\end{equation}
In addition, in the type IIA background, there is one unit of RR flux on the 
  $\P$ so that we have massless $\N=1$ $SU(N)$ vector multiplets arising 
from D2-branes wrapped on the vanishing cycles of the $A_{N-1}$ singularity 
at $z_{1,\ldots,4}=0$. Both from an M-theory and physical point of view, we 
know that  
there  are no massless scalars in the adjoint representation of the gauge 
group so that there is no 
analog of the Coulomb branch of the $\N=2$ $SU(N)$ theory. 
In type IIA, the blow up parameters of the $A_{N-1}$ singularity are
part of complex scalars, which are moduli in the $\N=2$ case. 
In presence of flux, there is a non trivial superpotential, whose
effect is to lift these flat directions. Classically, the blow down
geometry is then frozen and $SU(N)$ is not broken.

To see explicitly the transition from $\C_I$ to $\C_{I\!I}$, we perform the 
conifold transition from the resolved conifold to the deformed conifold. 
First of all, since $[\xi_1,\xi_2]$ are projective coordinates, they can't 
vanish simultaneously and we can replace the first Eq. of (\ref{resolved}) 
by the vanishing determinant of the system
\begin{equation}
(z_1+iz_2)(z_1-iz_2)+(z_3+iz_4)(z_3-iz_4)=0\, .
\label{conifold}
\end{equation} 
When the volume of the base $\P$ vanishes, the second  equation of 
(\ref{resolved}) that determines $\xi_{1,2}$ is then useless and we can omit 
it. Thus, we are left with Eq. (\ref{conifold}) that describes the conifold, 
which is singular at the origin 
$z_{1,\ldots,4}=0$.  The deformed conifold $T^* S^3$ is obtained by adding 
a constant  to the r.h.s.
\begin{equation}
z_1^2+z_2^2+z_3^2+z_4^2=\mu\, , 
\label{deformed}
\end{equation}
where $\mu$ can be chosen to be real 
and positive 
without loss of generality. 
The manifold $\C_{I\!I}$ is then obtained by taking into account the 
identification (\ref{ZN1}). To see explicitly that there is a lens space 
base $S^3/\Z_N$ in $\C_{I\!I}$, one identifies the $S^3$ in $T^* S^3$ as the 
set of points 
satisfying   Eq. (\ref{deformed}) with real $z_{1,\ldots,4}$ and rewrite Eq. 
(\ref{ZN1}) in the equivalent form
\begin{equation}
\begin{pmatrix} z_1 \\ z_2 \end{pmatrix} \equiv
\begin{pmatrix} \cos\left(\frac{2\pi}{N}\right) & 
~-\sin\left(\frac{2\pi}{N}\right) \\
\sin\left(\frac{2\pi}{N}\right) &
~\cos\left(\frac{2\pi}{N}\right) \end{pmatrix}
\begin{pmatrix} z_1 \\ z_2 \end{pmatrix}\, , \quad
\begin{pmatrix} z_3 \\ z_4 \end{pmatrix} \equiv
\begin{pmatrix} \cos\left(\frac{2\pi}{N}\right) & 
~-\sin\left(\frac{2\pi}{N}\right) \\
\sin\left(\frac{2\pi}{N}\right) &
~\cos\left(\frac{2\pi}{N}\right) \end{pmatrix}
\begin{pmatrix} z_3 \\ z_4 \end{pmatrix}\, .
\label{ZNlens}
\end{equation} 
As seen in Section \ref{KKreduc}, 
the full type IIA background in this phase contains a 
D6-brane wrapped on the base lens space. 
The D6-brane has the effect to break supersymmetry to $\N=1$ on its
world-volume. The closed string $U(1)$ vector multiplet associated to
the  base $\P$ of $\C_I$ 
is dual to the open string $U(1)$ on the world volume of the D6-brane. 

Finally, we would like to connect the moduli space of $\C_{I\!I\!I}$ to the 
two 
previous branches associated to $\C_I$ and $\C_{I\!I}$. The equations 
of the resolved conifold (\ref{resolved}) imply Eq. (\ref{conifold}) and the 
definition of the blow up $\P$ parametrized by $[\xi_1,\xi_2]$ at the origin 
was chosen to be 
\begin{equation}
\frac{z_1+iz_2}{z_3+iz_4}=-\frac{z_3-iz_4}{z_1-iz_2}=\frac{\xi_2}{\xi_1}\, .
\label{xi12}
\end{equation}
Performing the flop transition on $\P$ amounts to defining instead projective 
coordinates $[\zeta_1,\zeta_2]$ as
\begin{equation}
-\frac{z_1+iz_2}{z_3-iz_4}=\frac{z_3+iz_4}{z_1-iz_2}
=\frac{\zeta_2}{\zeta_1}\,  .
\label{zeta12}
\end{equation}
After the flop transition, the resolved conifold is then taking the form 
\cite{Candelas:1989ug}
\begin{equation}
\left\{
\begin{array}{l}
\phantom{-}(z_1+iz_2)\zeta_1+(z_3-iz_4)\zeta_2=0\\
-(z_3+iz_4)\zeta_1+(z_1-iz_2)\zeta_2=0
\end{array}
\right.\, .
\label{flop resolved}
\end{equation} 
To obtain a full definition of $\C_{I\!I\!I}$, we consider the $\Z_N$ 
orbifold on 
the variables $z_{1,\ldots,4}$ given in Eq. (\ref{ZN1})  and 
extend its action to $[\zeta_1,\zeta_2]$ such that it is a discrete isometry 
group of the resolved conifold  (\ref{flop resolved}),
\begin{equation}
\left\{
\begin{array}{l}
\mbox{\big (}(z_1\pm iz_2), (z_3\pm iz_4)\mbox{\big )}\equiv 
\left( e^{\pm 2\pi i/N}(z_1\pm iz_2),  
e^{\pm 2\pi i/N}(z_3\pm iz_4)\right)\\
{}[\zeta_1,\zeta_2]\equiv [e^{-2\pi i/N}\zeta_1,e^{2\pi i/N}\zeta_2]
\end{array}
\right. .
\label{flop ZN}
\end{equation}
The inhomogeneous coordinate on the chart $\zeta_1\neq 0$ is 
$\zeta_2/\zeta_1$ and the 
restriction of the orbifold action to the base $\P$ is 
$[1,\zeta_2/\zeta_1]\equiv [1,e^{4\pi i/N}\zeta_2/\zeta_1]$.  This 
corresponds to a $\Z_{N/2}$ action for $N$ even, and to a $\Z_N$ action 
for $N$ odd.
The fixed points on $\C_{I\!I\!I}$ satisfy $z_{1,\ldots,4}=0$, \ie are sitting 
on the base. 
For odd $N$, in the chart $\zeta_1\neq 0$, only the north pole  
$[1,0]$ is fixed, while in the chart $\zeta_2 \neq 0$, only the south
pole $[0,1]$ is fixed. For even $N$, not only these poles are fixed under
$\Z_N$, but the full base $\P$ is invariant under the
$\Z_2$ subgroup generated by the $(k=\frac{N}{2})$'th power of the above
generator. Thus, for 
any $N$, the peculiarities of the restriction of the $\Z_N$ action to the 
base together with the fixed points set  on $\C_{I\!I\!I}$ are precisely what 
was found in Appendix \ref{KKR} and summarized in Section \ref{KKreduc}.

\section{Linear $\sigma$-model descriptions}
\label{sigma}

In this section, we first give linear $\sigma$-model descriptions of 
the type IIA backgrounds $\C_{I}$ and $\C_{I\!I\!I}$, when the RR flux
is not included. Then, a way to take into account this flux is
actually to derive linear $\sigma$-model descriptions of the full $G_2$ 
holonomy spaces $\G_{I}$, $\G_{I\!I}$ and $\G_{I\!I\!I}$.

$\C_I$ and $\C_{I\!I\!I}$ are orbifolds of the resolved conifold. 
In order to describe them in terms of a linear $\sigma$-model, one
considers a two-dimensional $\N=(2,2)$ supersymmetric Abelian gauge theory 
whose low energy field configuration in a certain limit of parameter space
sweeps out the desired CY, \cite{Witten:1993yc}. 
For the model associated with the resolved conifold,
one introduces a $U(1)$ gauge field 
coupled to four chiral ones, whose charges are $Q=(1,-1,-1,1)$. Their scalar 
components $x_{1,\ldots,4}$ are thus subject to the $U(1)$ gauge equivalence  
\begin{equation}
\mbox{\big (}x_1,x_2,x_3,x_4\mbox{\big )}\equiv 
\left( e^{i\lambda}x_1,e^{-i\lambda}x_2,e^{-i\lambda}x_3, 
e^{i\lambda}x_4\right)\, ,
\label{u1}
\end{equation}
where $\lambda$ is real, and the vanishing D-term condition
\begin{equation}
|x_1|^2+|x_4|^2-|x_2|^2-|x_3|^2 = t\, , 
\label{D-term}
\end{equation}
where $t$ is a Fayet-Iliopoulos term.

To make contact with the description of Eqs. (\ref{resolved}) and 
(\ref{flop resolved}), we identify the $z_i$ variables with the gauge 
invariant combinations of the $x_i$'s under (\ref{u1})
\cite{Aganagic:2001ug}
\begin{equation}
z_1+iz_2 = x_1 x_2\, , \quad z_1-iz_2 = x_3 x_4\, , 
\quad -(z_3-iz_4)=x_1 x_3\, , \quad
z_3+iz_4=x_2x_4\, .
\end{equation}
Note that these relations are consistent with Eq. (\ref{conifold}). From Eq. 
(\ref{conifold}), one can see an explicit $U(1)\times U(1)$ isometry group 
acting as
\begin{equation}
(z_1\pm iz_2)\to e^{\pm i\alpha}(z_1\pm iz_2)\, , \qquad 
(z_3\pm iz_4)\to e^{\pm i\beta}(z_3\pm iz_4)\, ,
\end{equation}
that can be lifted to $\xi_{1,2}$ or $\zeta_{1,2}$ by consistency with Eqs. 
(\ref{xi12}) and (\ref{zeta12}). In terms of the original variables 
$x_{1,\ldots,4}$, this isometry group amounts to 
\begin{equation}
\mbox{\big (}x_1,x_2,x_3,x_4\mbox{\big )}\to 
\left(e^{i(\alpha-\beta)/2}x_1, e^{i(\alpha+\beta)/2}x_2, 
e^{-i(\alpha+\beta)/2}x_3, e^{-i(\alpha-\beta)/2}x_4\right)\, .
\end{equation} 
Now, the linear  $\sigma$-models of $\C_I$ and $\C_{I\!I\!I}$ are obtained by 
taking into account  the $\Z_N$ subgroup action  Eq. (\ref{ZN1}), whose 
generator corresponds to $\alpha=\beta=2\pi /N$:
\begin{equation}
\Z_N\,  : \quad  \mbox{\big (}x_1,x_2,x_3,x_4\mbox{\big )}\equiv
\left(x_1,e^{2\pi i/N} x_2,e^{-2\pi i/N} x_3,x_4\right) \,  ,
\label{ZN}
\end{equation}
which gives additional identifications between equivalence classes of
the $U(1)$ action (\ref{u1}).

For $t>0$ in Eq. (\ref{D-term}), one sees 
that $(x_1,x_4)\neq (0,0)$ and the whole space can be covered by two 
charts $x_1\neq0$ and $x_4\neq 0$. For $x_1\neq 0$, we can fix
representatives of the equivalence classes of (\ref{u1}) by using the
$U(1)$ action to remove the phase of $x_1$, \ie  in this chart 
representatives have the form $\left[r_1,x_2^{(1)},x_3^{(1)},x_4^{(1)}\right]$,
with $r_1>0$.
The $\Z_N$ action (\ref{ZN}) acts on these representatives as
\begin{equation}
\left[r_1,x_2^{(1)},x_3^{(1)},x_4^{(1)}\right]\equiv
\left[r_1,e^{2\pi i/N} x_2^{(1)},e^{-2\pi i/N} x_3^{(1)},x_4^{(1)}\right]\, .
\label{zn1}
\end{equation}
 The fixed points 
$\left[ \sqrt{t-|x_4^{(1)}|^2},0,0,x_4^{(1)}\right]$ 
parametrize the base $\P$ minus its north pole. Similarly, in the chart 
$x_4\neq 0$, fixing the representatives of the $U(1)$ equivalence classes 
to be of the form
$\left[x_1^{(4)},x_2^{(4)},x_3^{(4)},r_4\right]$ with $r_4>0$, the $\Z_N$ 
action 
looks analogous to (\ref{zn1}) and fixes the base $\P$ minus its south pole.
Altogether,  
there is a $\P$ of $A_{N-1}$ singularities in the $\sigma$-model description. 
Therefore, the phase $t> 0$ corresponds to $\C_I$. 

For $t< 0$ in Eq. (\ref{D-term}), one has instead $(x_2,x_3)\neq (0,0)$ 
and we can cover the space  with two charts $x_2\neq0$ and $x_3\neq 0$. 
For $x_2\neq 0$, fixing the representatives of the $U(1)$ classes to have 
the form $\left[x_1^{(2)},r_2,x_3^{(2)},x_4^{(2)}\right]$ with $r_2>0$, the
$\Z_N$ action (\ref{ZN}) amounts to
\begin{equation}
\left[x_1^{(2)},r_2,x_3^{(2)},x_4^{(2)}\right]\equiv
\left[e^{2\pi i/N} x_1^{(2)},r_2,e^{-4\pi i/N} x_3^{(2)},
e^{2\pi i/N}x_4^{(2)}\right]\, . 
\label{zn2}
\end{equation}
For $N$ even, we find again the $\Z_{N/2}$ action on the base $\P$ minus the 
north pole 
$\left[ 0,\sqrt{|t|-|x_3^{(2)}|^2},x_3^{(2)},0\right]$, while for 
$N$ odd, this action is of order $N$. For odd $N$, only the 
south pole $[0,\sqrt{|t|},0,0]$ is fixed under $\Z_N$, while for 
even $N$, in addition the whole $\P$ minus 
the north pole is fixed under a $\Z_2$ subgroup. 
Similarly, in the chart $x_3\neq 0$, one finds 
the same features, with the roles of the north and 
south poles inverted. After gluing together the two charts, the 
$\sigma$-model in the phase $t< 0$ appears thus to be associated  to 
$\C_{I\!I\!I}$. 

We would like now to consider the effect of adding flux. This can be
done by constructing $\sigma$-model descriptions of $\G_I$ and  
$\G_{I\!I\!I}$, the purely geometrical backgrounds which are dual in  
M-theory to the type IIA compactifications on $\C_I$ and
$\C_{I\!I\!I}$ in presence of RR flux. 
We shall use  the formalism introduced in \cite{Aganagic:2001ug}, where 
7-dimensional $G_2$ manifolds are described by linear $\sigma$-models.
However, we shall have in addition to implement 
the orbifold action in this formalism. 

As a starting point, let us recall the $\sigma$-model of $Spin(S^3)$ seen as 
a lift of the resolved conifold with one unit of RR flux through $S^2$. To 
this end, one considers  the original scalars $x_1$,\ldots,$x_4$ subject to 
the D-term constraint Eq. (\ref{D-term}), together with an additional 
$2\pi$-periodic real variable $\phi$, where the $U(1)$ modding action is 
now \cite{Aganagic:2001ug}
\begin{equation}
(x_1,x_2,x_3,x_4;\phi)\equiv 
(e^{i\lambda}x_1,e^{-i\lambda}x_2,e^{-i\lambda}x_3,e^{i\lambda}x_4;
\phi+\lambda) \, .
\label{U(1)M}
\end{equation}
Choosing  $\lambda=-\phi$, the 
representatives of the equivalence classes of this $U(1)$ action are taken to
be \mbox{$\left[x_1^{(0)},x_2^{(0)},x_3^{(0)},x_4^{(0)};0\right]$}.
The linear $\sigma$-model of $Spin(S^3)$ is then described in terms of the 
real D-term constraint on  the gauge fixed complex variables 
$x_i^{(0)}$,
\begin{equation}
|x_1^{(0)}|^2+|x_4^{(0)}|^2-|x_2^{(0)}|^2-|x_3^{(0)}|^2 = t\, ,
\label{Dtermlift}
\end{equation}
and the resulting space is thus 7-dimensional. For $t> 0$, it is an $\R^4$ 
fibration parametrized by $x_2^{(0)}$ and $x_3^{(0)}$, over an  $S^3$ base 
spanned by 
$x_1^{(0)}$ and $x_4^{(0)}$ at $x_2^{(0)}=x_3^{(0)}=0$. For $t< 0$, the roles 
of $(x_1^{(0)},x_4^{(0)})$ and $(x_2^{(0)}, x_3^{(0)})$ are reversed.  

To describe $\G_I$ and $\G_{I\!I\!I}$, we have to extend the $\Z_N$ action 
(\ref{ZN}) on $\phi$ as well. Let us consider a general anzatz
\begin{equation}
\Z_N\,  : \quad  \mbox{\big (}x_1,x_2,x_3,x_4;\phi \mbox{\big )}\equiv
\left(x_1,e^{2\pi i/N} x_2,e^{-2\pi i/N} x_3,x_4;
\phi+\kappa\right) \,  ,
\label{ZNlift}
\end{equation}
for some constant $\kappa$ that should be a multiple of $2\pi/N$. 
On our representatives of the $U(1)$ equivalence classes, this action amounts
to the identifications,
\begin{equation}
\left[x_1^{(0)},x_2^{(0)},x_3^{(0)},x_4^{(0)};0\right]\equiv
\left[e^{-i\kappa}x_1^{(0)},e^{i\kappa+2\pi i/N} x_2^{(0)},
e^{i\kappa-2\pi i/N} x_3^{(0)},e^{-i\kappa}x_4^{(0)};0\right]\, .
\label{znlift}
\end{equation}
Now, for $t>0$ associated to the lift to 7 dimensions of $\C_I$, we know 
from $\G_I$ that the orbifold action is trivial on the $S^3$ base. Thus, 
$\kappa = 0$,
\begin{equation}
\Z_N\, : \quad \left[x_1^{(0)},x_2^{(0)},x_3^{(0)},x_4^{(0)};0\right]\equiv
\left[x_1^{(0)},e^{2\pi i/N} x_2^{(0)},e^{-2\pi i/N} x_3^{(0)},x_4^{(0)};0
\right]\, ,
\label{znlift0}
\end{equation}
and we recognize the $\R^4/\Z_N$ fibration over $S^3$ of $\G_I$. Actually, to 
make contact with the definition (\ref{GI}), one should choose to 
parametrize $\R^4/\Z_N$ with $x_2^{(0)}$ and the complex conjugate 
$\bar{x}_3^{(0)}$ so that the $\Z_N$ is acting from the left. For 
$t< 0$, we have instead a lens space $S^3/\Z_N$ as expected for 
$\G_{I\!I\!I}$. 
Here also, to reproduce the $\Z_N$ left action convention in definition 
(\ref{GIII}), the lens space should be parametrized by $x_2^{(0)}$ and 
$\bar{x}_3^{(0)}$.

What we have seen is that the 2-cycle flop $t\to -t$ in the $\sigma$-models 
of $\C_I$ and $\C_{I\!I\!I}$ can be lifted to a 3-cycle flop in the 
7-dimensional 
$\sigma$-models of  $\G_I$ and $\G_{I\!I\!I}$. This point of view involves the 
$S^1\simeq U(1)_{2,L}$ fibration introduced in Section \ref{KKreduc}. 
As we 
reviewed there, $\G_I$ can also be obtained geometrically by lifting to 
M-theory the deformed conifold plus $N$ D6-branes wrapped on $S^3$, by 
introducing $S^1\simeq U(1)_{1,L}$. However, since there is no linear  
$\sigma$-model description of the deformed conifold, this point of view 
cannot help to derive a linear $\sigma$-model description of $\G_I$. 

On the other hand,
to see explicitly in a $\sigma$-model language the 3-cycle flop between 
$\G_{I\!I}$ and $\G_{I\!I\!I}$, it is instead the $S^1\simeq U(1)_{1,L}$ 
fibration 
over the resolved conifold with $N$ units of RR flux through the $S^2$ that 
can be used  \cite{Aganagic:2001ug}. One starts from the linear $\sigma$-model 
of the resolved conifold, Eqs. (\ref{u1}) and (\ref{D-term}), where the $N$ 
units of flux are on the $S^2$ parametrized by $x_{1}$ and $x_4$ for 
$t> 0$, while for $t< 0$, 
the flux is through the flopped $S^2$ 
parametrized by $x_2$ and $x_3$. The lift to the $G_2$ holonomy manifolds  
$\G_{I\!I}$ and $\G_{I\!I\!I}$ is done by adding to the D-term condition 
(\ref{D-term}) the $2\pi$-periodic variable $\phi'$ and replacing the 
modding action (\ref{u1}) by the identification
\begin{equation}
\mbox{\big (}
x_1,x_2,x_3,x_4;\phi'
\mbox{\big )}
\equiv 
\left(e^{i\lambda}x_1,e^{-i\lambda}x_2, e^{-i\lambda}x_3, e^{i\lambda}x_4;
\phi'+N \lambda\right) \, .
\label{u1m2}
\end{equation}
Gauging away $\phi'$ amounts to choosing $\lambda =-\phi'/N +2k\pi/N$ 
with any integer $k$ in the range $k\in\{0,\ldots,N-1\}$. 
Representatives of these equivalence
classes can thus be chosen to be of the form 
$\left[{x'_1}^{\!(0)}\! ,{x'_2}^{\!(0)}\! ,{x'_3}^{\!(0)}\! ,{x'_4}^{\!(0)};0
\right]$, 
subject to the remaining discrete $\Z_N$ identification
\begin{equation}
\Z_N \, : \ 
\left[{x'_1}^{\!(0)}\! ,{x'_2}^{\!(0)}\! ,{x'_3}^{\!(0)}\! ,{x'_4}^{\!(0)};0
\right]
\equiv \left[e^{2\pi i/N}{x'_1}^{\!(0)}\! ,e^{-2\pi i/N}{x'_2}^{\!(0)}\! ,
e^{-2\pi i/N}{x'_3}^{\!(0)}\! ,e^{2\pi i/N}{x'_4}^{\!(0)};0\right]  ,
\label{znlift02}
\end{equation}
and the D-term condition 
\begin{equation}
|{x'_1}^{\!(0)}|^2+|{x'_4}^{\!(0)}|^2-|{x'_2}^{\!(0)}|^2-|{x'_3}^{\!(0)}|^2 
= t\, .
\label{Dtermlift2}
\end{equation}
We choose to associate 
the phases $t> 0$ to $\G_{I\!I}$ and $t< 0$ to $\G_{I\!I\!I}$. We can make 
contact with the definition (\ref{GIII2}) of $\G_{I\!I\!I}$ by noting that the 
$\Z_N$ acts on the right both on the $S^3$ spanned by ${x'_2}^{\!(0)}$ and 
$\bar{\phantom{x}} \;\!\!\!\!\! {x'_3}^{\!(0)}$ and on $\R^4$ parametrized 
by $\bar{\phantom{x}} \;\!\!\!\!\! {x'_1}^{\!(0)}$ and ${x'_4}^{\!(0)}$. 

We now have two different descriptions of $\G_{I\!I\!I}$, the first one in 
Eqs. 
(\ref{Dtermlift}) and (\ref{znlift0}) for $t< 0$, and the second in Eqs. 
(\ref{Dtermlift2}) and (\ref{znlift02}) for $t< 0$. To see  explicitly 
that these descriptions are equivalent, 
we want to give the coordinate transformations that map them into each other.
In the first description, $\G_{I\!I\!I}$ is covered 
by two charts ${x_2}^{(0)}\neq 0$ and ${x_3}^{(0)}\neq 0$, due to Eq. 
(\ref{Dtermlift}). When ${x_2}^{(0)}\neq 0$, the phase 
$\arg ({x_2}^{(0)})$ is well defined, 
so that we can introduce new coordinates 
${x'_j}^{\!(0)} =x_j^{(0)}e^{i n_j \arg ({x_2}^{(0)}) }$ with 
$(n_1,\ldots,n_4)=(1,-2,0,1)$. The $\Z_N$ action (\ref{znlift0})
on the old coordinates then implies indeed the action (\ref{znlift02}) on
the new ones. The inverse coordinate change is
$x_j^{(0)}={x'_j}^{\!(0)} e^{i n_j \arg ({x'_2}^{\!(0)})}$.
In the chart ${x_3}^{(0)}\neq 0$,  the phase 
$\arg ({x_3}^{(0)})$ is well defined and the analogous coordinate
transformations are
${x'_j}^{\!(0)}=x_j^{(0)}e^{-i n_j \arg ({x_3}^{(0)})}$, whose  inverse is
$x_j^{(0)}={x'_j}^{\!(0)} e^{i n_j \arg ({x'_3}^{\!(0)})}$. 

Similar invertible coordinate transformations that map the $\Z_N$ actions
into each other don't exist for $t> 0$, as it should, since in this case
the two models are associated to the different spaces $\G_I$ and 
$\G_{I\!I}$.

\vskip 1cm
\noindent
{\bf Acknowledgments:}

\vskip .15in

\noindent
We are grateful to I. Antoniadis, I. Bakas, I. Brunner, P. Mayr and
S. Theisen for useful discussions. 
We acknowledge the CERN theory division, where part of 
this work has been done. While at Ecole Polytechnique, P.K.\ was
supported by an European Commission Marie Curie postdoctoral
fellowship under the contract number HPMF-CT-2000-00919.
The work of H.P. is supported in part by the European 
networks HPRN-CT-2000-00148, MRTN-CT-2004-503369, MRTN-CT-2004-005104
and the E.C. Excellence Grant MEXT-CT-2003-509661.

\appendix

\vspace{1cm}
\begin{center}
{\bf \Large Appendices}
\end{center}

\section{The $G_2$ holonomy metrics}
\label{metrics}

In this appendix we review the construction of $G_2$ metrics on 
$Spin(S^3)$, the idea of which we have already recalled around Eq. 
(\ref{s3s3}). We discuss their isometry groups and in particular the action
of their subgroups (\ref{KKu1}) along which we want to perform 
KK reductions.

We represent $g\in SU(2)\simeq S^3$ either as
\begin{equation}
g=\begin{pmatrix} Z &i Z' \\ i\bZ' & \bZ \end{pmatrix}
\, ,\quad\mbox{with}\quad
Z,Z'\in\Co
\quad\mbox{and}\quad
\det(g)=|Z|^2+|Z'|^2=1\, ,
\end{equation}
or introduce the Euler angles $\theta \in (0,\pi)$, $\varphi\in[0,2\pi)$ and
$\psi\in[0,4\pi)$ as coordinates on the open complement of the set 
$\{Z=0\}\cup \{Z'=0\}$,
\begin{equation}
\begin{array}{lll}
g &\! =\!\exp\left(\frac{i}{2}\psi \si_3\right)
\exp\left(\frac{i}{2}\theta \si_1\right)
\exp\left(\frac{i}{2}\varphi \si_3\right) 
 &\!\! =\!\!\begin{pmatrix} 
\!\!\!\cos(\theta/2)e^{i(\psi+\varphi)/2} 
&\!\!\! i\sin(\theta/2)e^{i(\psi-\varphi)/2} \\
i\sin(\theta/2)e^{-i(\psi-\varphi)/2} 
&\!\!\! \,\phantom{i}\cos(\theta/2)e^{-i(\psi+\varphi)/2} 
\end{pmatrix}  ,
\end{array}
\label{euler}
\end{equation}
where
\[
\si_1=\begin{pmatrix} 0 & 1 \\ 1 & 0 \end{pmatrix}\ , \qquad 
\si_2=\begin{pmatrix} 0 & -i \\ i & 0 \end{pmatrix}\ , \qquad 
\si_3=\begin{pmatrix} 1 & 0 \\ 0 & -1 \end{pmatrix}
\]
denote the Pauli matrices. 
Define
\begin{equation}
\Theta_j:=g_j^{-1}dg_j=\frac{i}{2}\omega^a_{j,L}\si_a\, , \qquad (j=1,2,3)
\end{equation}
where $\omega^a_{j,L}$ are the three left-invariant one-forms on the $j$-th
copy of $SU(2)$ in (\ref{s3s3}),
\begin{align}
&w_{j,L}^1=
\cos(\varphi_j)d\theta_j+\sin(\varphi_j)\sin(\theta_j)d\psi_j\, , \nonumber\\
&w_{j,L}^2=
\sin(\varphi_j)d\theta_j-\cos(\varphi_j)\sin(\theta_j)d\psi_j\, , \\
&w_{j,L}^1= d\varphi_j+\cos(\theta_j)d\psi_j \, .\nonumber
\end{align}
In \cite{Atiyah:2001qf} the $G_2$ metric is given
in variables
\begin{align}
da^2&:=-{\rm Tr}\left(\Theta_2\otimes\Theta_2+\Theta_3\otimes\Theta_3
-\Theta_2\otimes\Theta_3-\Theta_3\otimes\Theta_2\right)\, , \nonumber\\
db^2&:=-{\rm Tr}\left(\Theta_1\otimes\Theta_1+\Theta_3\otimes\Theta_3
-\Theta_1\otimes\Theta_3-\Theta_3\otimes\Theta_1\right)\, , \\
dc^2&:=-{\rm Tr}\left(\Theta_1\otimes\Theta_1+\Theta_2\otimes\Theta_2
-\Theta_1\otimes\Theta_2-\Theta_2\otimes\Theta_1\right)\nonumber
\end{align}
as
\begin{equation}
ds^2=\frac{dr\otimes dr}{1-(r_0/r)^3}+\frac{r^2\left(1-(r_0/r)^3\right)}{72}
\left(2da^2-db^2+2dc^2\right) +\frac{r^2}{24}db^2\, , \label{met1}
\end{equation}
where $r\in[r_0,\infty)$, with $r_0$ being a modulus of the solution 
describing the radius of the base $S^3$ of $Spin(S^3)$. The three group
elements $g_j$ are subject to the identification of Eq.\ (\ref{s3s3}),
which we can use e.g.\ to fix $g_3\equiv 1$, bringing the above metric to
the form,
\begin{equation}
ds^2=\frac{dr\otimes dr}{1-(r_0/r)^3}
-\frac{r^2\left(1-(r_0/r)^3\right)}{72}
{\rm Tr}\left[(2\Theta_2-\Theta_1)\otimes (2\Theta_2-\Theta_1)\right]
-\frac{r^2}{24}{\rm Tr}\left(\Theta_1\otimes \Theta_1\right)\, .\label{met2}
\end{equation}
Since
\[
-{\rm Tr}\left(\Theta_j\otimes \Theta_k\right)
=\frac{1}{2}\sum_{a=1}^3 \left(\omega^a_{j,L}\otimes \omega^a_{k,L}\right)
\, ,
\]
where in particular
\begin{align}
-\frac{1}{2}{\rm Tr}    &\left(\Theta_1\otimes \Theta_1\right)
=\frac{1}{4}\sum_{a=1}^3 \left(\omega^a_{1,L}\otimes \omega^a_{1,L}\right)
\nonumber\\
&=\frac{1}{4}\left(\rule{0mm}{4mm}d\theta_1\otimes d\theta_1
+d\varphi_1\otimes d\varphi_1
+d\psi_1\otimes d\psi_1 +\cos(\theta_1)[d\varphi_1\otimes d\psi_1
+d\psi_1\otimes d\varphi_1]
\right) \nonumber \\
&=\frac{1}{2}\left.
\left(dZ_1 \otimes d\bZ_1 + d\bZ_1\otimes dZ_1 + dZ_1' \otimes d\bZ_1' 
+ d\bZ_1'\otimes dZ_1'\right)
\right|_{\{|Z_1|^2+|Z_1'|^2=1\}}
\end{align}
is the standard metric on the three-sphere inherited from its embedding into
Euclidean $\R^4$, the metric (\ref{met2}) indeed
agrees with the complete $G_2$ holonomy metric 
originally given in \cite{Gibbons:1989er}. 
In this example, the sphere parametrized
by $g_1$ becomes the base of the fibration, whereas the one associated to
$g_2$ is ``filled'' and turned into $\R^4$. Thus, the metric (\ref{met2}) 
corresponds to the manifold we call  $\G_{I\!I}$ in the text, once we take 
into account the relevant $\Z_N$ isometry orbifold action.  

Note that the $SU(2)_{1,L}\times SU(2)_{2,L}\times SU(2)_{3,L}$ isometry
group (\ref{iso1}) of (\ref{met1}),
after gauge fixing $g_3\equiv 1$, is identified with  an 
$SU(2)_{1,L}\times SU(2)_{2,L}\times [SU(2)_{1,R}\times SU(2)_{2,R}]_D$
isometry group of (\ref{met2}), where the subindex $D$ denotes the diagonal
subgroup, 
\begin{equation}
\begin{array}{lcl}
 & & (g_1,g_2,1)\to (h_1g_1h_3^{-1},h_2g_2h_3^{-1},1)\, ,\\
\mbox{where}&&(h_1,h_2,h_3)
\in SU(2)_{1,L}\times SU(2)_{2,L}\times SU(2)_{3,L}\, .
\end{array} 
\label{iso2}
\end{equation}
We are particularly interested in the action of  $U(1)$ isometry subgroups 
of some $SU(2)_{j,L}$  represented by matrix multiplication of 
\begin{equation}
\xi(\omega)=\begin{pmatrix} e^{i\omega} & 0 \\ 0 & e^{-i\omega} \end{pmatrix}
\in SU(2)_{j,L}\, .
\end{equation}
On the one hand, the left-action $m_L(\xi(\omega),g)=\xi(\omega) g$ acts as
\begin{equation}
m_L\: : \left\{ \begin{array}{rcl}
(Z,Z') &\mapsto & (e^{i\omega}Z,e^{i\omega}Z') \\
{}& \mbox{or} & {} \\ 
(\theta,\varphi,\psi) & \mapsto & (\theta,\varphi,\psi+2\omega)\, , \\
\end{array} \right.
\label{mL}
\end{equation}
where we can periodically continue the range of the Euler angle $\psi$, 
\ie define it
modulo $4\pi$. On the other hand, the right-action 
$m_R(\xi(\omega),g)=g \xi(\omega)$ acts as
\begin{equation}
m_R \: : \left\{ \begin{array}{rcl}
(Z,Z') &\mapsto & (e^{i\omega}Z,e^{-i\omega}Z') \\
{}& \mbox{or} & {} \\ 
(\theta,\varphi,\psi) & \mapsto &\left\{
\begin{array}
{l@{\, ,\, }l@{\, ,\, }l@{)\ ,\quad \mbox{for}\ }c@{\ \leq \omega <\ }c}
(\theta & \varphi+2\omega      & \psi      & 0       & \pi-\varphi/2\, , \\
(\theta & \varphi+2\omega-2\pi & \psi+2\pi & \pi-\varphi/2  
& 2\pi-\varphi/2\, , \\
(\theta & \varphi+2\omega-4\pi & \psi      & 2\pi-\varphi/2 & 2\pi\, . \\
\end{array} \right. \\
\end{array} \right. \label{mR}
\end{equation} 
To obtain $\G_{I\!I}$, one considers the modding action of the $\Z_N$ subgroup
of $SU(2)_{1,L}$, which thus identifies $\psi_1\equiv
\psi_1+4\pi/N$. The base in  $\G_{I\!I}$ is then the lens space we
shall denote  $L_N:=S^3/m_L(\Z_N)$.
The metrics on $\G_{I}$ and $\G_{I\!I\!I}$ are constructed analogously.

\section{Kaluza-Klein reduction of the lens space $L_N:=S^3/m_L(\Z_N)$}
\label{KKlens}

We want to recall in this appendix how the Kaluza-Klein reduction of
$\G_{I\!I}$ and $\G_{I\!I\!I}$ along the orbit of the isometry $U(1)_{1,L}$
results in a geometry that is a fibration over $S^2$, with $N$ units of
RR two-form flux through $S^2$. This amounts to concentrate in
particular on the Hopf reduction of the lens space $L_N:=S^3/m_L(\Z_N)$. 

In general, when performing the Kaluza-Klein reduction of an
eleven-dimensional metric in M-theory (in the Einstein frame)
\[
d\hat{s}^2=\sum_{\hat\mu,\hat\nu=0}^{10}\hat{g}_{\hat\mu\hat\nu}\, 
dx^{\hat\mu}\otimes dx^{\hat\nu}\, ,
\]
where $x^{10}\in[0,2\pi)$ parametrizes the orbit of a $U(1)$ isometry, one
identifies the metric 
$ds^2=\sum_{\mu,\nu=0}^{9}g_{\mu\nu}\, dx^\mu\otimes dx^\nu$, 
RR one-form gauge potential $A=A_\mu dx^\mu$ and dilaton $\Phi$ in type IIA 
via
\begin{equation}
\hat{g}_{\hat\mu\hat\nu}=\begin{pmatrix}
e^{-2\Phi/3}g_{\mu\nu} +e^{4\Phi/3} A_\mu A_\nu 
& ~~e^{4\Phi/3} A_\mu \\ 
e^{4\Phi/3} A_\nu & ~~e^{4\Phi/3} 
\end{pmatrix}\, . \label{KK}
\end{equation}
Assuming that $g_{\mu\nu}$, $A_\mu$ and $\Phi$ do not depend on $x^{10}$, the 
$11$-dimensional Einstein-Hilbert action reduces to the $10$-dimensional
Einstein-Hilbert-Maxwell action 
(in the string frame). 
The gauge potential $A$ obtained this way
may be well-defined only upon performing a gauge transformation, but its 
curvature $F=dA$ will in general be well-defined. 

$\G_{I\!I}$ and $\G_{I\!I\!I}$ are $\R^4$-fibrations over 
$L_N:=S^3/m_L(\Z_N)$, where the base can be identified with the
$SU(2)$ parametrized by $g_1$ in Eqs. (\ref{GII}) and
(\ref{GIII}) and modded by the left-action of its $\Z_N$ subgroup. 
When seeing this $SU(2)\simeq S^3$ itself as a Hopf fibration, $H\, :\,
S^3\to S^2$, over a two-sphere spanned by $(\theta_1,\varphi_1)$ via
\begin{align}
&X_1 = 2\, {\rm Re}(Z_1 \bZ_1')=Z_1 \bZ_1'+\bZ_1 Z_1'
=\sin(\theta_1)\cos(\varphi_1)\, , \nonumber \\
&X_1' = 2\, {\rm Im}(Z_1 \bZ_1')=-i(Z_1 \bZ_1'-\bZ_1 Z_1')
=\sin(\theta_1)\sin(\varphi_1)\, , \label{hopf} \\
&X_1'' = |Z_1 |^2 -|Z_1' |^2=Z_1 \bZ_1-Z_1' \bZ_1'
=\cos(\theta_1) \ , \nonumber
\end{align}
the fiber parametrized by $\psi_1$ is generated by the orbit of the
left-action (\ref{mL}) of $U(1)_{1,L}$ on $S^3\simeq SU(2)$, the base
being invariant under this action.
Hence $L_N$ and thus $\G_{I\!I}$ and $\G_{I\!I\!I}$ are fibrations over
$S^2$. 

In order to determine the RR two-form flux through the $S^2$, we can restrict
to the base part
\begin{align*}
& d\hat{s}^2= 
-\frac{R^2}{24}{\rm Tr}\left(\Theta_1\otimes \Theta_1\right) \\
&\!\!=\frac{R^2}{48} \!\left(\! d\theta_1\otimes d\theta_1 \!
+ \! d\varphi_1\otimes d\varphi_1 \!
+ \! \frac{4}{N^2}d\tpsi_1\otimes d\tpsi_1 \! 
+ \! \frac{2}{N}\cos(\theta_1)[d\varphi_1\otimes d\tpsi_1 \! 
+ \! d\tpsi_1\otimes d\varphi_1]\right)
\label{dhs3}
\end{align*}
of (\ref{met2}),
where $\tpsi_1=N\psi_1/2\in[0,2\pi)$ and 
$R^2=r^2\left(1+\frac{1-(r_0/r)^3}{3}\right)$. With (\ref{KK}) we read off,
\begin{equation}
e^{4\Phi/3} = \frac{R^2}{12 N^2}\, , \;
ds^2= \frac{R^3}{96\sqrt{3}N}
\left(d\theta_1\otimes d\theta_1+\sin^2(\theta_1)d\varphi_1\otimes
d\varphi_1 \right) \, , \;
A = \frac{N}{2}\cos(\theta_1)d\varphi_1\, .
\end{equation}
From the curvature
\[
F=dA=-\frac{N}{2}\sin(\theta_1)\, d\theta_1\wedge d\varphi_1\, ,
\]
one computes as first Chern class of the bundle 
$H'\, :\, L_N\to S^2$ defined as in Eq. (\ref{hopf}),
\begin{equation}
c_1(H')=\frac{1}{2\pi}\int_{S^2}F=-\frac{N}{4\pi}
\int_{\varphi_1=0}^{2\pi}\int_{\theta_1=0}^{\pi}\sin(\theta_1)\, 
d\theta_1\wedge d\varphi_1
=-N\, ,
\label{Nflux}
\end{equation}
which also gives the units of RR two-form flux through the $S^2$.

\section{Kaluza-Klein reduction of $(S^3\times \R^4)/m_R(\Z_N)$}
\label{KKR}

In this appendix, we want to perform the Kaluza-Klein reduction of 
$\G_{I\!I\!I}$ along the orbits of $U(1)_{2,L}$. In the case where we have
used the identification (\ref{s3s3}) to fix the gauge $g_1\equiv 1$ in
(\ref{GIII2}), the orbifold action turns into a simultaneous {\em
right}-action of $\Z_N$ on $g_2$ and $g_3$, as explained in Appendix
\ref{metrics}. 
This action is free due to the noncontractibility of the $S^3$
parametrized by $g_2$.  
The metric on $\G_{I\!I\!I}$ looks like (\ref{met2}) with 
$\Theta_1 \mapsto \Theta_2$ and $\Theta_2\mapsto \Theta_3$. 
In addition, we identify points that are mapped into each other 
by the simultaneous right-action   
action (\ref{mR}) on $g_2$ and $g_3$ for $\omega=2\pi k/N$, with 
$k=0,\ldots,N-1$.
In order to analyze the consequences for the 
Kaluza-Klein reduction, we distinguish the cases of odd respectively
even $N$.

\paragraph{$\mathbf{N}$ odd:}

Consider first generic points such that $Z_2,Z_2'\neq 0$, so that we can
use Euler angles $\theta_2,\varphi_2,\psi_2$ on the noncontractible
three-sphere, whereas we use complex coordinates $Z_3,Z_3'$ on the
$\R^4$-fiber. 
Let $\varphi_2\in[0,2\pi/N)$. Then
the $\Z_N$ right action (\ref{mR}) identifies the points 
\begin{equation}
(\theta_2,\varphi_2,\psi_2,Z_3,Z_3') \! \equiv \!
\mbox{\big (}\theta_2,\varphi_2+\frac{2\pi k}{N},
\psi_2+2\pi k\ {\rm mod}\ 4\pi,
(-1)^k e^{i\pi k/N}Z_3,(-1)^k e^{-i\pi k/N}Z_3'\mbox{\big )} ,
\end{equation}
for $k=0,\ldots,N-1$. This action is always free on the noncontractible
$S^3$ para\-metrized by $(\theta_2,\varphi_2,\psi_2)$. 

Performing the KK reduction (\ref{KK}) 
[with $\tpsi_2=\psi_2/2\in[0,2\pi)$] means integrating over the orbits 
of the isometry $\psi_2\to \psi_2+2\omega$ with $\psi_2$ defined modulo 
$4\pi$. 
On the resulting six-dimensional space, we have to identify the points 
\begin{equation}
(\theta_2,\varphi_2,Z_3,Z_3')\equiv
\mbox{\big (}\theta_2,\varphi_2+\frac{2\pi k}{N},
(-1)^k e^{i\pi k/N}Z_3,(-1)^k e^{-i\pi k/N}Z_3'\mbox{\big )}\, ,
\end{equation}
where $k=0,\ldots,N-1$ and again $\varphi_2$ is taken to lie in the range 
$[0,2\pi/N)$.
This results in a space of topology
\begin{equation}
\left(S^2\times \R^4\right)/\Z_{N}\, ,
\end{equation}
where the $\Z_{N}$ acts on the two-sphere by rotations around the axis
through the north and south poles by angles $2\pi k/N$, combined with an
action
\[
(Z_3,Z_3')\equiv\left( (-1)^k e^{i\pi k/N}Z_3,
(-1)^k e^{-i\pi k/N}Z_3'\right)\, 
\]
on $\R^4$,
for $k=0,\ldots,N-1$, which generates the familiar action (\ref{mR}) for
$\Z_N$ on $\R^4$ when running through all values of $k$.
Note that on the six-dimensional space, the orbifold action now has 
two fixed points, namely the poles of the base $S^2$, where
either $Z_2=Z_3=Z_3'=0$ or $Z_2'=Z_3=Z_3'=0$. Since we had previously
excluded these loci, let's discuss them now.

Consider the five-dimensional sublocus $\{Z_2'=0\}\simeq S^1\times \R^4$ 
inside $\G_{I\! I\! I}$ and let $Z_2=e^{i\tilde\varphi_2}$. 
The $\Z_N$ orbifold acts on it as   
\begin{equation}
(\tilde\varphi_2,Z_3,Z_3')\equiv
\mbox{\big (}\tilde\varphi_2+\frac{2\pi k}{N},
e^{2\pi i k/N}Z_3,e^{-2\pi ik/N}Z_3'\mbox{\big )}\, , \label{poles}
\end{equation}
where $k=0,\ldots,N-1$. The KK reduction is along the $S^1$ parametrized
by $\tilde\varphi_2$ resulting in the sublocus
\{south pole\}$\times \R^4$ with the familiar $\Z_N$ action on 
$\R^4$.
The discussion of the sublocus $\{Z_2=0\}$ is completely analogous upon
replacing $\tilde\varphi_2$ by $-\tilde\varphi_2'$ and the south pole by the
north pole.  

Over each point on the six-dimensional space, there was a $U(1)$-orbit that 
gives rise to a $U(1)$ gauge connection $A$. Over a generic
point, the orbifold group doesn't act within the gauge fiber over this point
-- it merely gives a relative half-twist (depending on $k$ mod 2)
between the disjoint orbits over its $N$ pre-image points. Over the
fixed points, \ie the north and south poles of the $S^2$ (times the origin
in $\R^4$), on the contrary, 
the orbifold group $\Z_N$ acts within the gauge fibers over these
two points. 
We interpret the behavior at generic points as indicating one
unit of RR two-form flux through the   $S^2$, since
performing the reduction (\ref{KK}) for 
the $G_2$ metric on $\G_{I\!I\!I}$ [with $\tpsi_2=\psi_2/2\in[0,2\pi)$] 
leads to
\begin{align*}
e^{4\Phi/3}&=\frac{1}{36r}(4r^3-r_0^3)\; ,\\
A&=\frac{1}{2}\cos(\theta_2)d\varphi_2
-\frac{r^3-r_0^3}{(4r^3-r_0^3)}
\left( \rule{0mm}{4mm} \cos(\theta_2)d\varphi_3
+\sin(\theta_2)\sin(\varphi_2-\varphi_3)d\theta_3\right. \\
&\hspace*{40mm}\left. \rule{0mm}{4mm} 
+[\cos(\theta_2)\cos(\theta_3)+\cos(\varphi_2-\varphi_3)\sin(\theta_2)
\sin(\theta_3)]d\psi_3\right)
\end{align*}
and gives one unit of RR two-form flux through the covering $S^2$ parametrized
by $\varphi_2\in[0,2\pi)$ and $\theta_2\in(0,\pi)$,
\[
\frac{1}{2\pi}\int_{S^2} dA =-1\, .
\]
The additional
$\Z_N$ action within the $U(1)$ gauge fiber over the poles is interpreted
as shifting the masses of string states that arise in the twisted sectors
associated to these fixed points. This mechanism was introduced in 
\cite{Schwarz:1995bj} in order to describe the type IIA dual of the 
six-dimensional CHL compactification. It may also be thought of
as additional flux localized at the fixed points. 

\paragraph{$\mathbf{N}$ even:}

Again we consider first generic points such that $Z_2,Z_2'\neq 0$,
and let $k=l+\frac{N}{2}\alpha$, with 
$l=0,\ldots,N/2-1$ and $\alpha=0,1$. For
$\varphi_2 \in[0,4\pi/N)$, the $\Z_N$ right action (\ref{mR}) identifies 
all the points with
coordinates
\begin{equation}
(\theta_2,\varphi_2,\psi_2,Z_3,Z_3')\!\equiv\!
\mbox{\big (}\theta_2,\varphi_2+\frac{2\pi l}{N/2},
\psi_2+2\pi\alpha \ {\rm mod}\ 4\pi,
(-1)^{\alpha} e^{2\pi i l/N} Z_3,
(-1)^{\alpha} e^{-2\pi i l/N} Z_3'\mbox{\big )}, 
\end{equation}
for $l=0,\ldots,N/2-1$ and $\alpha=0,1$. Again this action is always free on 
the noncontractible $S^3$ parametrized by $(\theta_2,\varphi_2,\psi_2)$.

Performing the KK reduction (\ref{KK}) 
[with $\tpsi_2=\psi_2/2\in[0,2\pi)$] means integrating over the orbits 
of the isometry $\psi_2\to \psi_2+2\omega$, with $\psi_2$ defined modulo 
$4\pi$. 
On the resulting six-dimensional space, we have to identify the points
\begin{equation}
(\theta_2,\varphi_2,Z_3,Z_3')\equiv
\mbox{\big (}\theta_2,\varphi_2+\frac{2\pi l}{N/2},
(-1)^{\alpha} e^{2\pi i l/N} Z_3,
(-1)^{\alpha} e^{-2\pi i l/N} Z_3'\mbox{\big )}\, ,
\end{equation}
where $l=0,\ldots,N/2-1$ and $\alpha=0,1$ and again $\varphi_2$ is 
taken to lie in the range $[0,4\pi/N)$.
This results in a space of topology
\begin{equation}
\left(S^2\times \R^4\right)/(\Z_{2}\times\Z_{N/2})\, ,
\end{equation}
where the $\Z_{2}$ acts trivially on the two-sphere 
and the $\Z_{N/2}$ by rotations around the axis
through the north and south poles by angles $4\pi l/N$, whereas the
$\Z_2\times\Z_{N/2}$ acts on the $\R^4$ as
\[
(Z_3,Z_3')\equiv\left( e^{2i\pi (l+\frac{N}{2}\alpha)/N}Z_3,
e^{-2i\pi (l+\frac{N}{2}\alpha)/N}Z_3'\right)
=\left( e^{2i\pi k/N}Z_3,
e^{-2i\pi k/N}Z_3'\right)\, ,
\]
with $k=l+\frac{N}{2}\alpha$, where $l=0,\ldots,\frac{N}{2}-1$ 
and $\alpha=0,1$.
Note that the generator of $\Z_2$ always leads to an 
$S^2/\Z_{N/2}$ of $A_1$ singularities given by $Z_3=Z_3'=0$.
The poles of the base, where in addition either $Z_2=0$ or $Z_2'=0$, 
are fixed by the whole $\Z_N$. 
The discussion of these two fixed points and their $\R^4$ fibers is
word for word  
the same as the one around (\ref{poles}) for odd $N$. 

Over a generic point of the six-dimensional space, 
the orbifold group doesn't act within the $U(1)$ gauge fiber over this
point 
-- it merely gives a relative half-twist (depending on $\alpha$)
between the disjoint orbits over its $N$ pre-image points. Over the
$\Z_2$-fixed base $S^2/\Z_{N/2}$, however, these
half-twists generate a $\Z_2$ action within the gauge fiber over it.
At the $\Z_N$-fixed points (the poles of the $S^2$ times the origin in $\R^4$) 
there is a localized 
embedding of the
$\Z_N$-action 
within the gauge fiber over them. 
By the same calculation as for odd $N$, we interpret
the behavior at generic points as indicating one
unit of RR two-form flux through the $S^2$ parametrized by 
$(\theta_2,\varphi_2)$. The additional
$\Z_2$ (respectively $\Z_N$) action localized within the $U(1)$ gauge fiber 
over the $\Z_2$ (respectively $\Z_N$)-fixed locus 
implies that these fixed sets don't lead to new massless states through their
associated twisted sectors. In particular there are no massless 
vector bosons associated to the vanishing cycle of the $A_1$ singularity
at the $\Z_2$-fixed sphere.

This mechanism of removing massless states in type IIA string theory
from the twisted sectors
associated to the fixed point set of an orbifold action by 
a localized embedding of 
this action also in the $U(1)$ 
gauge bundle of 
the RR one-form 
potential is exactly what was also used in \cite{Schwarz:1995bj} to 
construct a type IIA dual of the six-dimensional CHL compactification.
It would be interesting to better understand this mechanism in the type IIA
setting but unfortunately we don't know the CFT description of this
nontrivial RR-background. It is, however, very reminiscent of a familiar
shift orbifold on an additional circle describing the gauge bundle.


\end{document}